\documentclass{article}
\usepackage{graphics}
\usepackage{amssymb}
\def\PrPC{{PrPC}}

\begin{document}

\title{Integrity of H1 helix in prion protein revealed by 
molecular dynamic simulations to be especially vulnerable to 
changes in the relative orientation of H1 and its S1 flank
\thanks{This paper has been accepted for pubcliation in European Biophysical Journal on Feb 2, 2009.}}





\author{Chih-Yuan Tseng$^1$\thanks{\emph{Present address:} Department of Oncology, University of Alberta, Edmonton, AB T6G 1Z2 Canada. E-mail: %
chih-yuan.tseng@ualberta.ca; richard617@gmail.com}, Chun-Ping Yu$^1$ and HC Lee$^{1,2}$\\
$^1$Department of Physics and $^2$Graduate Institute of Systems Biology and 
Bioinformatics\\
National Central University, Chungli, Taiwan 320}
\date{}

\maketitle

\begin{abstract}

In the template-assistance model, normal prion protein (PrPC), the
pathogenic cause of prion diseases such as Creutzfeldt-Jakob (CJD) in
human, Bovine Spongiform Encephalopathy (BSE) in cow, and scrapie in
sheep, converts to infectious prion (PrPSc) through an autocatalytic
process triggered by a transient interaction between PrPC and
PrPSc. Conventional studies suggest the S1-H1-S2 region in PrPC to be
the template of S1-S2 $\beta$-sheet in PrPSc, and the conformational
conversion of PrPC into PrPSc may involve an unfolding of H1 in PrPC
and its refolding into the $\beta$-sheet in PrPSc. Here we conduct a
series of simulation experiments to test the idea of transient
interaction of the template-assistance model.  We find that the
integrity of H1 in PrPC is vulnerable to a transient interaction that
alters the native dihedral angles at residue Asn$^{143}$, which
connects the S1 flank to H1, but not to interactions that alter the
internal structure of the S1 flank, nor to those that alter the
relative orientation between H1 and the S2 flank.

\end{abstract}


\section{Introduction}

Prion protein (PrP) in its infectious form is the pathogen that causes
several prion diseases such as Creutzfeldt-Jakob (CJD) in human,
Bovine Spongiform Encephalopathy (BSE) in cow, and scrapie in sheep
\cite{Prusiner98}. Two reviews recently summarize past studies and discuss molecular mechanisms of 
the prion disease to understand physiological function of PrP and 
pathogenic pathways \cite{Aguzzi08a,Aguzzi08b}.

\begin{figure}[ht]
\centering
\resizebox{0.75\textwidth}{!}{\includegraphics{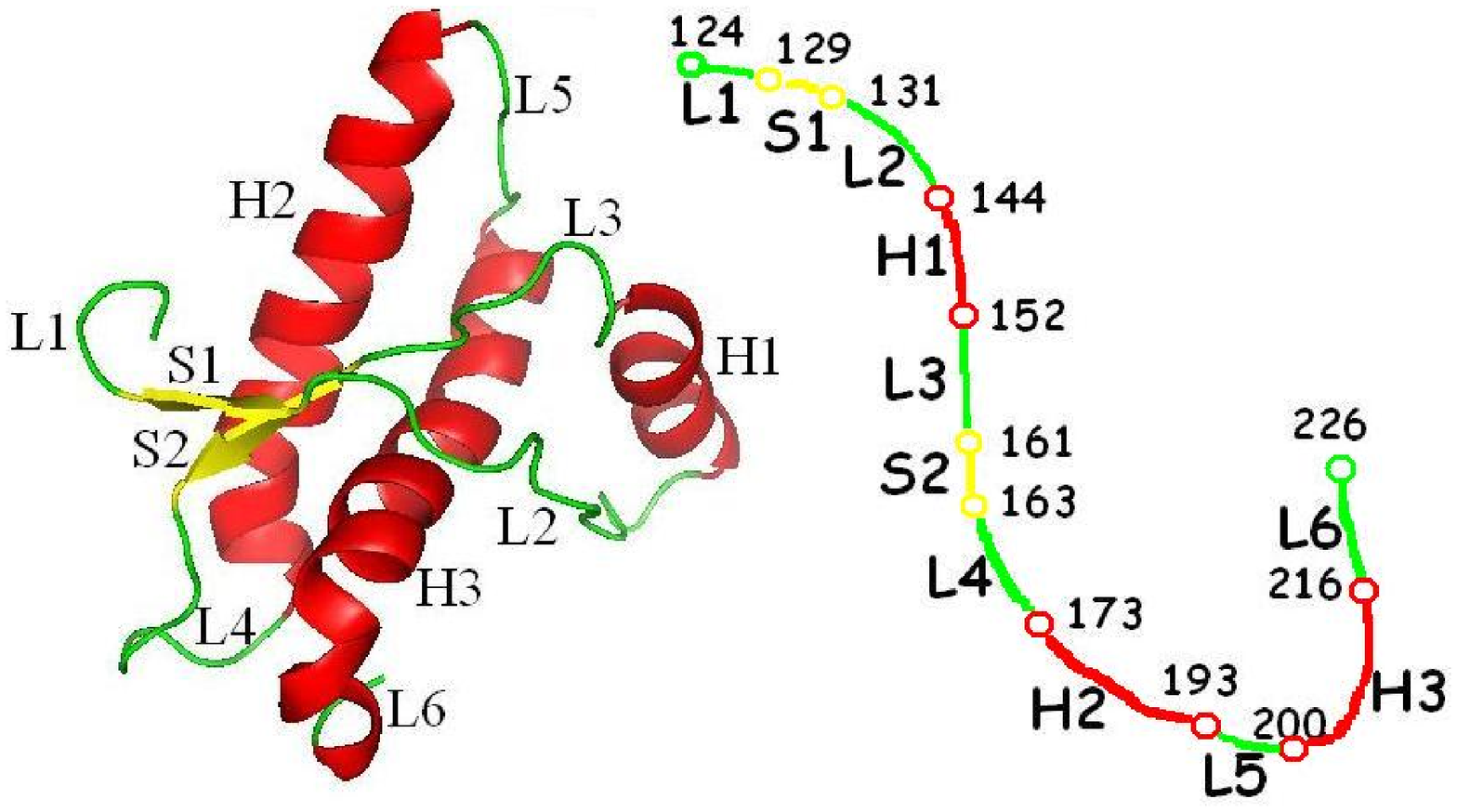}}
\caption{}
\label{f:1ag2-motif}
\end{figure}

An \textit{in vitro} experiment conducted by \cite{Castilla06} provides 
strong evidence for the protein-only hypothesis. 
Fig. \ref{f:1ag2-motif} shows an NMR structure of
the C-terminal of mouse PrP in its native form (\PrPC) (PDB code:
1AG2). It contains 103 residues from Gly$^{124}$ to Tyr$^{226}$
classified into secondary structures and surface loops
\cite{Riek96}. These include three $\alpha$-helices: H1, residues 144
to 152, H2 (173-193), and H3 (200-216); two $\beta$-strands: S1
(129-131) and S2 (161-163), which form an anti-parallel $\beta$-sheet;
six loops: L1 (124-128), L2 (132-143), L3 (153-160), L4 (164-172), L5
(194-198), and L6 (217-226).  In what follows we shall refer to the
S1-L2 segment as the S1 flank, or F1, the L3-S2 segment as the S2
flank (F2), and the segment from S1 to S2, inclusive, as the S1-H1-S2
peptide.  Homologues of PrP in other organisms generally have residue
numbering that differ from the mouse numbering given in
Fig. \ref{f:1ag2-motif}; unless explicitly otherwise specified, in
this text we will use the mouse numbering.
\marginpar{Fig. 1 is here.}

Experimental investigations suggest that the pathogeny of prion
diseases is characterized by the unfolding of PrPC followed by
misfolding into the infectious scrapie isoform (PrPSc) \cite{Pan93} ,
which involves conformational changes in the C-terminal residues
121-231 but no chemical reaction \cite{Harris99,Jackson00}.  Muramoto
et al. showed the H2 and H3 helices seem to be stabilized by disulfide
bonds and likely to have the same conformation in PrPC and PrPSc
\cite{Muramoto96}. They also found the deletion of both H2 and H3 from
PrPC does not stop its conversion to PrPSc.  Eghiaian et al. showed
the epitope of antibody to be conserved in the H2 and H3 regions,
again suggesting that these two helices are conserved during
conversion \cite{Eghiaian04}.

Because the H2-H3 region seems to be conserved during the PrPC to
PrPSc conversion, recent experimental and computational investigations
have focused on the S1-H1-S2 region (residue 124 to 163), and these have revealed
important features of the prion disease. 
A predominantly helical propensity in
the H1 region is demonstrated by several \textit{in silico} studies under different 
experimental conditions by \cite{Santini03} and \cite{Santini04}.
On the question of 
possible mechanisms that trigger the conformational conversion, \textit{
in vitro} studies suggest that altering the pH level of the solvent,
which varies static electric interactions, might destabilize the H1
helix and trigger conversion \cite{Zou02,Calzolai03}.  Similar
effects were observed in \textit{in silico} studies by \cite{Levy01}, \cite{Levy02} and by \cite{DeMarco04}.  In
simulations on mouse prion 1AG2, Guilbert et al. found that a major
modification of dihedral angles around the residue 125 was required
for the formation of a $\beta$-strand in residues 121-133
\cite{Guilbert00}. In addition, DeMarco and Daggett show 
detachment of the S1-H1-S2 region 
from the rigid H2-H3 region to accommodate misfolding
and the H1 helix is only partially 
unfolded \cite{DeMarco07}. 
By studying the isolated S1-H1-S2 peptide, Sharman et al. and Ziegler et al. 
both 
showed a predominantly helical propensity in the H1 region for PrP in water as well 
using CD and NMR methods \cite{Sharman98,Zeigler03}. 
This property is further 
demonstrated by Dima and Thirumalai's \cite{Dima04} and by Watzlawik et al.'s 
studies \cite{Watzlawik06}. When Watzlawik et al. experimentally clarified role 
of the H1 helix in the conversion, they found the H1 helix does not convert into $\beta$-sheet 
yet it promotes aggregation. In addition, Kozin et
al. found that residues 142-166 in human numbering (same numbering as
in mouse) has the propensity to form a $\beta$-hairpin around residues
153-156 at neutral pH level as well \cite{Kozin01}, which they believe to be
the event that drives the conversion.  
In addition, Derreumaux showed the residues 127-164 region 
has two equi-energetic conformations with $\beta$ or $\alpha$ features \cite{Derreumaux01}. 
Megy et al. provide an experimental evidence for this structural duality \cite{Megy04}.
They found the $\beta$ structure in the region spanning 
residues from 142 to 167 very similar to NMR spectroscopy $\beta$ structure. 
These results may be summarized as follows: in the
PrPC to PrPSc conversion the S1-H1-S2 region plays an important role while the 
H2-H3 region plays at most a passive role, and during the conversion the $\beta$-sheet 
breaking in PrPC is likely to be the first barrier \cite{Barducci06} and 
detachment of the S1-H1-S2 region from the rigid H2-H3 region is another barrier \cite{DeMarco07}.
The H1 helix is only partially unfolded. Thus, here we only aimed to study H1 stability
using the S1-H1-S2 peptide.

One of the models advanced for the pathogenesis of the prion disease
is the template-assistance model \cite{Prusiner98,Horiuchi99,Tompa02}.
In this model it is assumed that \PrPC, normally more stable than
PrPSc in isolation, would in the presence of PrPSc convert to the
latter via a transient catalytic interaction with it.  The implication
is that a dimer of PrPSc's is energetically more stable than a system
of non-interacting PrPC and PrPSc, which was confirmed by Morrissey 
and Shakhnovich's computational analyses \cite{Morrissey99}.  
When there are other PrPC present,
the initial autocatalytic process would then lead to a propagation of
PrPC to PrPSc conversion.  Because PrPSc's always appear in
aggregated state and not in isolation, its structure is not precisely
known at present \cite{DeMarco04,Kozin01}, rendering an investigation
of the conversion-causing transient interaction between PrPC and PrPSc
problematic.  Nevertheless, an in-principle feasibility of the
template-assistance model was demonstrated by Malolepsza et
al. \cite{Malolepsza05}.  In their computer simulations the PrPSc was
approximated by peptides with $\beta$-sheet structure, and the authors
found that such peptides were able to induce conversion of peptides
with $\alpha $-helix.

A key assumption of the template-assistance model is that the the PrPC
to PrPSc conversion is triggered by a transient interaction, as
opposed to, say, a series of slow-acting contacts.  Here, we use
computer simulation to explore possible consequences of transient
interactions that may trigger the PrPC to PrPSc conversion, without
explicitly including the latter in the simulation.  In practice, we
investigate what sudden changes to the conformation of PrPC would
destabilize its native structure. 
If a conversion triggering transient interaction is found,
then a possible next step is to see whether (in simulation) the
presence of a PrPSc in the vicinity of a PrPC indeed would affect such
an interaction.  We take a two-step approach because a general
exploration of possible transient interactions between PrPC and PrPSc
by MD simulation would exceed our present computational capability,
and because an accurate knowledge of the conformation of PrPSc is
lacking.

As mentioned in \cite{Aguzzi08a}, a clear understanding of the physiological function of the PrPC and its interaction partner 
is still lacking, and this in turn prevents us from understanding the roles 
PrPC and PrPSc play in the pathogenesis of prion diseases.
Thus, in the present work we institute structural changes in PrPC \textit{ by
hand}, changes that we assume may be caused by hypothetical transient
interactions, and follow the aftermath in each case by molecular
dynamics (MD) simulation. Our goal is to investigate the dynamics of 
structural 
modifications of PrPC under simulations. Specifically we focus on the stability of
H1 after a change made in the S1-H1-S2 peptide (residues 124-167).  Our study contains
two parts. In the first part we attempt to determine changes in which
one of the two flanks - F1 and F2 - is more likely to initiate an
unfolding of H1.  We find the answer to be F1.  In the second part, we
initiate specific structural changes in F1, run MD simulations on the
S1-H1-S2 peptide, and focus our attention on the way H1 is affected.
We find the native structure of the S1-H1-S2 peptide to be generally
quite robust. Among structural alterations made to F1, modification of
the two dihedral angles of Asn$^{143}$ is found to be most likely to
lead to the unfolding of H1, and that when H1 unfolds, it tends to
form a $\beta$-hairpin turn at residues 150-152, which is close to the
153-156 region reported in \cite{Kozin01}.  Our results also suggest
that hydrophobic forces do not play a major role in the conversion
process.

\section{Method}
\label{method}
\subsection{Simulation parameter settings}
\label{settings}
The simulation package AMBER 8 \cite{amber} is used for energy
minimization and MD simulation with the AMBER force field
ff03. As a prelude to each simulation a full
conjugate gradient energy minimization is applied for 1000 iterations
to allow the spatial positions of the atoms to relax to their
respective local energy minima.  During minimization and MD simulation
SHAKE \cite{Ryckaert97} is invoked to
constrain hydrogen bonds. This has the effect of preventing the fast
bond vibration motion of hydrogens. 
The cut-off distance for non-bonded interactions
in all calculations is set to be 15 Angstrom.

For the MD simulation, system temperature is at room
temperature, or 300 K. Andersen temperature coupling is
employed to regulate temperature between protein and the
environment. 
Simulation time step is 2 fs. Initial velocities of atoms in proteins are
generated from Boltzmann distributions at 300 K temperature.  The
effect of solvent is represented by the modified generalized Born
model of Onufriev et al. \cite{Onufriev00}, where the pH is set
to neutral, and where, for calculating the effective Born radius, the
maximum distance between a pair of atoms is set to be 12 Angstrom. 
Periodic boundary 
condition is not applied in the calculations. 
The cut-off distance of non-bonded interactions 
is set to be 15 Angstrom.

The MD simulations are performed on a 32-node PC cluster at the
National Center for High-performance Computing in Taiwan.  Under the
above settings the average CPU time needed to simulate the folding of 
a 40-amino-acid peptide for 1 ns is about half an hour.

\subsection{Designs for two series of simulations on S1-H1-S2}
\label{twoexperiments}
Our intention is to represent the effect of potential 
conversion-triggering transient interactions on the S1-H1-S2 peptide (residues 124 to 167) by
artificially induced structural changes to the peptide, and study the
stability of the affected peptide through MD simulation. Following
this strategy we carry out two series of exploratory simulations
classified according to the artificial changes made to the peptide and
designated E$n$f$m$, where $n$= 1 and 2 is the classifier, and $m$
enumerates the simulations in each class.  In practice, a specific
designation indicates a specific initial conformation for the
peptide. In addition, f0 (without the prefix E$n$) denotes the
benchmark simulation in which the S1-H1-S2 peptide has the native
conformation as its initial conformation.

\noindent \textbf{The E1 series.}
The goal of this series of simulations is to identify which flank, 
F1 or F2, plays a crucial role in the stability of H1.  We consider three
extreme cases.  In each case, either F1, F2, or both together, is
completely pruned from the S1-H1-S2 peptide and the remainder,
initially in its respective native conformation, is simulated.  In
E1f1, F2 is pruned and the remainder consists of residues 124 to 154.
In E1f2, F1 is pruned and the remainder consists of residues 142 to
167. In E1f3, both flanks are pruned and the remainder consists of
residues 142 to 154.  When F1 is pruned, native contacts of 
mouse PrPC including the hydrogen bonds in the anti-parallel $\beta$-sheet, 
between the residue Arg$^{136}$ and the four residues Met$^{154}$, Tyr$^{155}$, Tyr$^{157}$, 
 and Pro$^{158}$, and between residues Pro$^{137}$ and Tyr$^{150}$ are broken. 
In contrast no native contacts other than the hydrogen bonds 
in the anti-parallel $\beta$-sheet are broken when F2 is pruned. 
It indicates F1 plays a more important role in the stability of H1 than F2. 
This notion is discussed further in section \ref{resultsadiss}.  
The total simulation time for this series is
approximately 800 ns.  

\begin{figure}[ht]
\centering
\resizebox{1\textwidth}{!}{\includegraphics{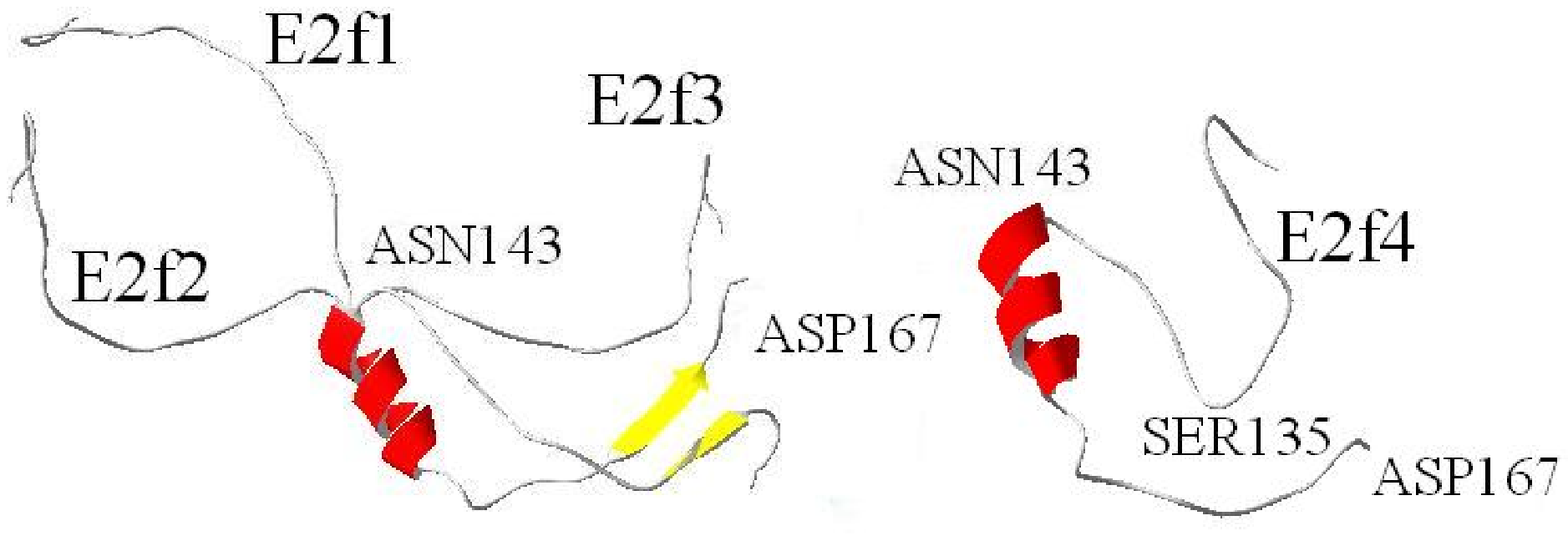}}
\caption{}
\label{f:initial_E2f}
\end{figure}

\noindent \textbf{The E2 series.}
Guilbert et al. \cite{Guilbert00} pointed out that a major
modification of the dihedral angles of a residue in F1 is required for
the formation of a $\beta$-sheet on the S1-H1-S2 peptide.  This,
together with the result from the E1 simulations, mainly that F2 plays a minor role in the stability of H1 relative to F1, motivate the E2
simulations describe below.  In each simulation, the initial
conformation of F2 relative to H1 is unchanged, and one of two types
of changes is made on F1.  In the first type, the dihedral angles of
Asn$^{143}$, the residue joining F1 to H1, are changed.  The native
values of the dihedral angles are $\Phi_0=-124.854^{o}$ and
$\Psi_0=132.794^{o}$, which lie within the $\beta$-strand region in
the Ramachandran plot.  In three simulations, designated E2f1, E2f2
and E2f3, the initial values of ($\Phi$, $\Psi$) are changed to
($\Phi$, $\Psi$)$_1$=(60$^{o}$, 60$^{o}$), ($\Phi$,
$\Psi$)$_2$=($\Phi_0$, 0$^{o}$), and ($\Phi$, $\Psi$)$_3$=(-60$^{o}$,
$\Psi_0$), respectively, see Fig.\ref{f:initial_E2f}.
\marginpar{Fig. 2 is here.}
These modifications are made using DeepView/Swiss-Pdb viewer
\cite{deepview}. Fig.\ref{f:initial_E2f}, as is
Fig.\ref{f:1ag2-motif}, are generated by Pymol \cite{pymol}.  In the
Ramachandran plot, ($\Phi$, $\Psi$)$_1$ lies in the left-handed
helical region, ($\Phi$, $\Psi$)$_2$ in the right-handed helical
region, and ($\Phi$, $\Psi$)$_3$ in the $\beta$-strand region.  The
total time for the three simulations is about 1 $\mu$s.  In all initial conformations 
in the above E2 series runs, native contacts including the hydrogen bonds in 
the anti-parallel $\beta$-sheet, between the residue Arg$^{136}$ and the four 
residues Met$^{154}$, Tyr$^{155}$, Tyr$^{157}$, Pro$^{158}$, and between residues 
Pro$^{137}$ and Tyr$^{150}$ are broken. No new contacts are formed.

In the second type, for which only one simulation (E2f4) is run, the initial
internal conformation of F1 is altered by changing the values of
dihedral angles of Ser$^{135}$ from $\Phi$=-75.575$^{o}$ and
$\Psi$=150.389$^{o}$ to $\Phi$=-75$^{o}$ and $\Psi$=60$^{o}$.  This
change has the effect of causing residues 124-143 to form a
$\beta$-hairpin-like structure (right panel in 
Fig.\ref{f:initial_E2f}). In this case, only the hydrogen bonds in the 
anti-parallel $\beta$-sheet 
(in native mouse PrPC) are broken.

\subsection{In-house analysis tool}
We employ several in-house tools to analyze the simulation 
results. The code PTRAJ from the AMBER package is used to 
extract peptide conformation at 1 ns intervals, 
and the program DSSP \cite{dssp} to
identify protein secondary structure.   
Two programs, g\_sas and g\_saltbr, from the MD package GROMACS \cite{GROMACS,GROMACS2,GROMACS3} are used to
calculate solvent accessible surfaces (SAS) and salt-bridge distances,
respectively. We will use a similarity matrix to define a distance between two simulations. 
Confidence interval of the similarity thereafter 
is estimate using the maximum likelihood estimate.

\section{Results}
\label{resultsadiss}

\subsection{Results of the f0 simulation}

\begin{figure}[ht]
\centering
\resizebox{1\textwidth}{!}{\includegraphics{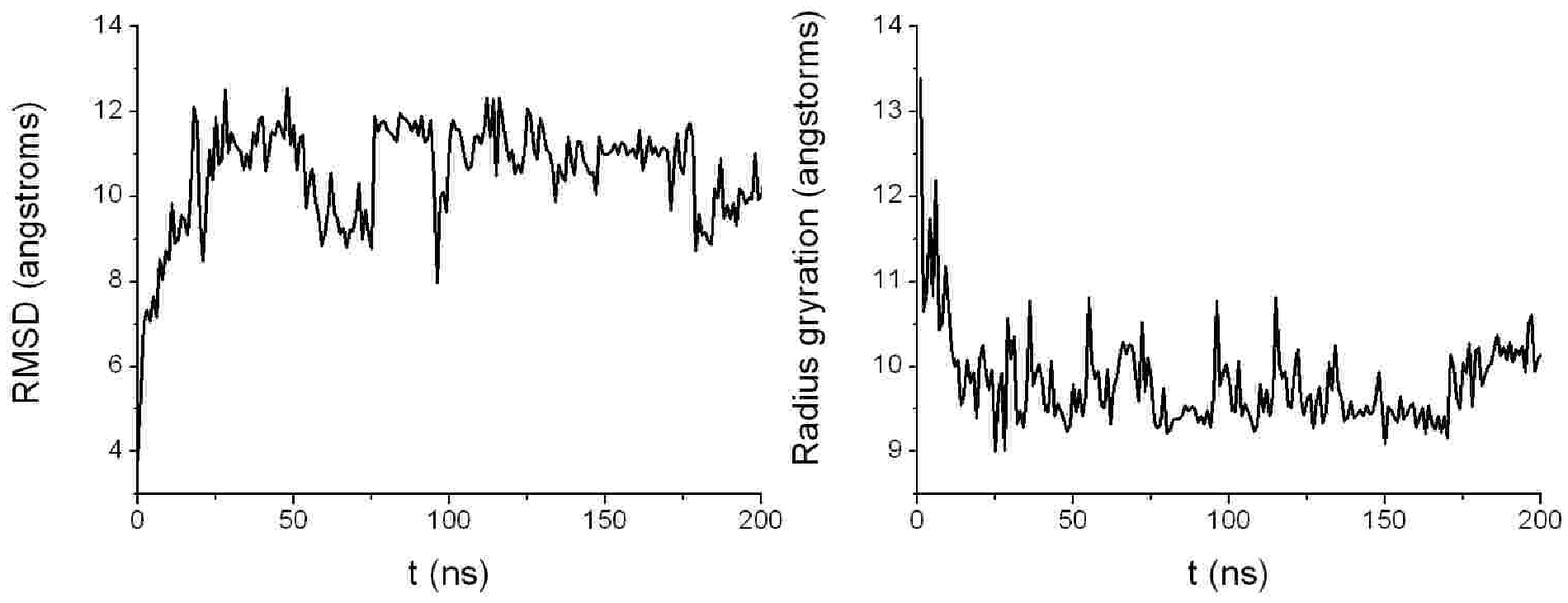}}
\caption{}
\label{rmsdargyr}
\end{figure}

\noindent \textbf{Stability of the S1-H1-S2 peptide.}
The benchmark experiment f0, was simulated for 200 ns during 
which, at intervals of 1 ns, the conformation 
of the S1-H1-S2 peptide is extracted. 
In the simulation of S1-H1-S2 (in mouse), the 
S1-H1-S2 forms a stable conformation different from its native state characterized by two 
quantities, the root mean square deviation (RMSD) and radius of gyration, shown in Fig.\ref{rmsdargyr}, 
but H1 remains largely intact. The RMSD calculates 
$C_\alpha$ atoms's 
positional difference between conformation in the simulation and the native conformation. 

Although the RMSD in the left panel of Fig. \ref{rmsdargyr} shows two drops around 50 and 175 
ns, the 
average RMSD is around 11 Angstrom during the whole simulation. 
The radius of gyration indicates spatial 
extent of all $C_\alpha$ atoms.  
The right panel of Fig. \ref{rmsdargyr} shows that the spatial extent of 
the structure 
remains stable after collapsing around 12 ns. One may attribute the collapse to the following 
reason. Barducci et al.'s studies suggest Tyr$^{128}$, Arg$^{164}$, and Asp$^{178}$ 
stabilize the $\beta$-structure in the S1-H1-S2 peptide \cite{Barducci06}. Particularly, mutation of Asp$^{178}$ 
will severely cause disruption of the $\beta$-sheet. Thus one should expect unzipping of the 
$\beta$-sheet in 
our simulation because of absence of the Asp$^{178}$ in the initial conformation. As shown in Fig. \ref{f:conf_tran_E2} 
in next section there is the disruption of the $\beta$-sheet after 5 ns. In addition, 
there is a stable H-bonded turn around Met$^{138}$ formed after 1 ns for the whole simulation. These two 
reasons may result in the collapse of conformation. 

\marginpar{Fig. 3 is here.}

\begin{figure}[ht]
\centering
\resizebox{1\textwidth}{!}{
\includegraphics{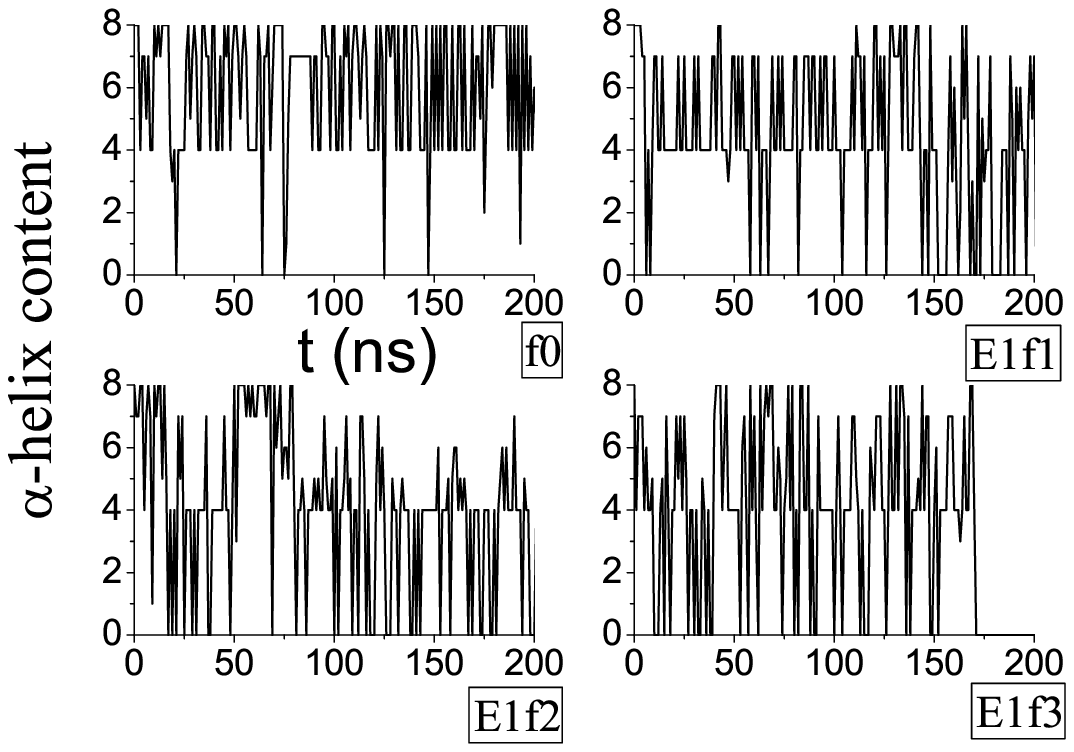}}
\caption{}
\label{f:H1-stability}
\end{figure}

\noindent \textbf{Helical propensity in H1.}
 The number of the residues from the H1 region 
- D$^{144}$WEDRYYR$^{151}$ - forming 
the current $\alpha$-helix is recorded at intervals of 
1 ns in the 200 ns simulation. 
The result of the content variation along the 
simulation time is 
plotted in f0 panel of Fig. \ref{f:H1-stability}. 
Since it takes 3.6 residues to form a turn in 
an $\alpha$-helix, one can define
a zero-turn helix that consists of 0 to 2 residues, 
one full turn 
helix contains residue 3 to 5, and two full turns helix contains 6 to 8 residues. 
This coarse-grained 
description provides a simple way to account for 
helical structural variations. In the f0 simulation 
H1 has zero turn in 8 accumulative ns 
out of a total 200 ns, one full turn in 70 out of 200 ns, and 
two full turns in 122 out of 200 ns. There is a strong helical propensity 
in H1. These results are consistent with expectations \cite{Dima04}.

\marginpar{Fig. 4 is here.}


\subsection{Results of E1 simulations}
\label{rd1stexp} 
\noindent \textbf{Stability of H1 is more dependent on F1.} The number of residues
forming the $\alpha$-helix in H1 region in E1f1, E1f2, and E1f3 is recorded.  
Each of the three simulations were run for 200ns. In each case, various data, including 
the number of residues forming the alpha-helix in ¡§H1 region¡¨ (residue 144 to 152) 
were recollected at 50 ps intervals. The plots in Fig. \ref{f:H1-stability} show the number of residues 
at every 20th data points, or effectively, data taken at 1 ns intervals. We will 
then apply a similarity analysis in the following to analyze the simulation results.

\begin{figure}[ht]
\centering

\resizebox{1\textwidth}{!}{\includegraphics{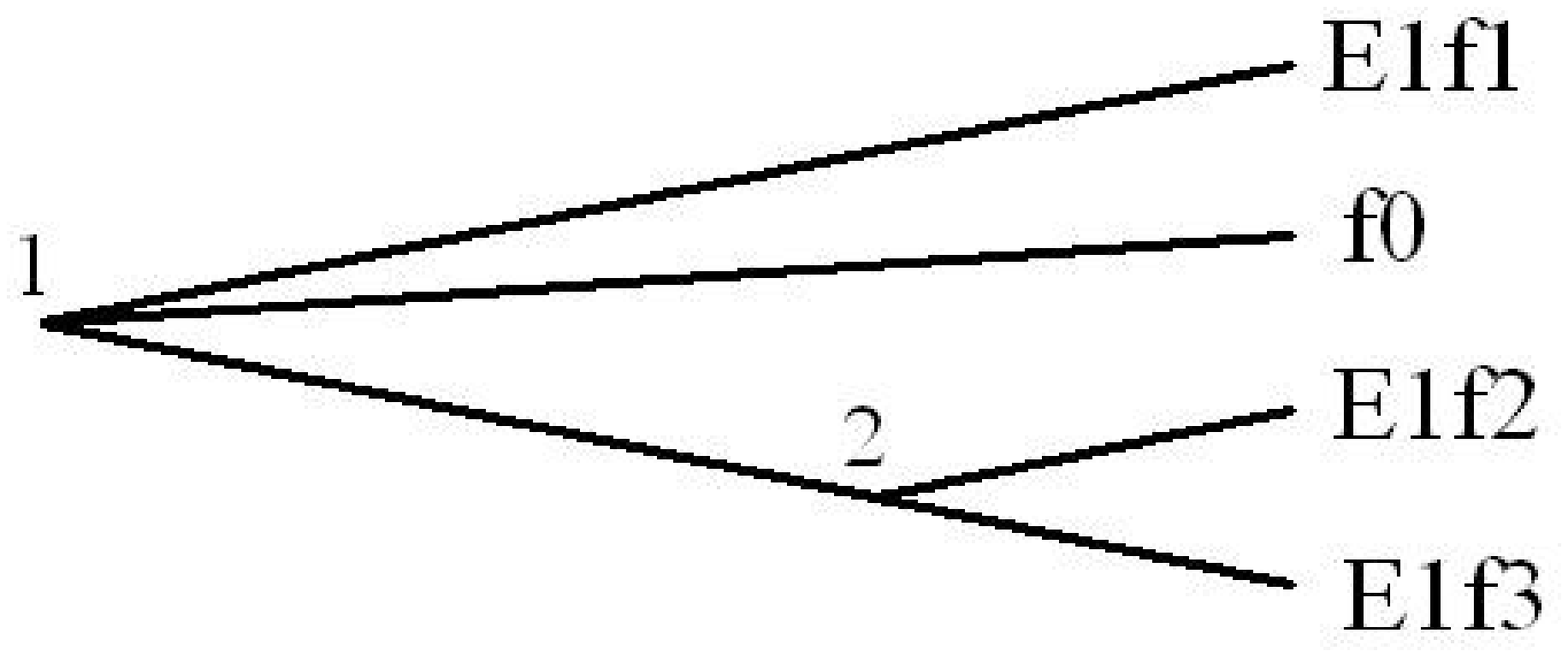}}
\caption{}
\label{f:E1-family tree}
\end{figure}

First, similarity of the f0 and the E1 simulations 
is defined as inner-product of 
two unit vectors, 
$s$. A vector characterized the simulation contains 
nine elements, which are accumulative 
times in 200 ns simulation of 0, 1,..., 8-residue in H1 forming 
a $\alpha$-helix correspondingly. Then the unit vector is obtained by normalizing the vector by its length. 
The value of $s$ ranges from 0 to 1. A zero inner product indicates the two vectors to be completely different 
and one denotes the two to be identical. Thus one obtains a four by four similarity matrix to reveal all correlations 
among the f0 and three E1 simulations $\it{S}$. Thereafter, we define 
a distance matrix, $\it{D}=1-\it{S}$ to reveal similarities among the simulations. Fig. \ref{f:E1-family tree} shows an example of 
a schematic plot of family tree. Notes that each end point of branch in the tree 
represents a simulation and then 
total branch's length between two end points denotes 
their differences. For example length between the f0 and the 
E1f1 $l_{f0-E1f1}=
l_{f0-1}+l_{1-E1f1}$. Longer the branch's length is, the two simulations differ more. 
After twenty family trees corresponding to data recorded at different intervals are  
constructed, we apply the maximum 
likelihood estimate to evaluate confidence interval of 
length between two simulations. 
The 95 \% confidence interval 
of length from the f0 to the E1f1 is $l_{f0-E1f1}=0.1192 \pm 0.0114$ unit length. 
The lengths between 
the f0 and two other simulations the E1f2 and the E1f3 are $l_{f0-E1f2}=0.172 
\pm 0.0067$ and $l_{f0-E1f1}=0.1875 \pm 0.0102$ 
respectively.

The lengths calculations indicate the $l_{f0-E1f1}$ to be the shortest at 
confidence level 0.95, which indicates E1f1
to be  most similar to f0, and the E1f2 and E1f3 are 
less similar to f0. Recall that in the E1f1 simulation F1 is retained in the 
peptide, 
in the E1f2 simulation F2 is retained, and in the E1f3 
simulation neither is retained, we thus conclude that relative to F2, F1 is
significantly more crucial to the stability of the $\alpha$-helix
nature of H1 and, by inference, that a transient interaction altering
the structure of F1 is more likely to lead to a PrPC to PrPSc 
conversion than a transient interaction altering the structure of F2. 
\marginpar{Fig. 5 is here.}

\subsection{Results of E2 simulations}
\label{rs2ndexp}
\smallskip\noindent \textbf{Native conformation of S1-H1-S2 has 
the lowest energy in simulation.}
Peptides in the E2 series have the same amino acid sequence but have
different initial conformation. The total
energy difference $\Delta E$(f$n$) = $E$(E2f$n$)-$E$(f0) = 41.7, 52.7, 37.9 and 23.2 kcal/mole for 
$n$= 1, 2, 3 and 4 respectively, where for every case the energy is taken after the initial
energy minimization and before the simulation begins.
These results confirm the expectation that the native conformation 
of the S1-H1-S2 peptide has the lowest energy, at least compared 
to the initial conformations imposed on the peptide in the E2 series of 
simulations. This also provides a minimum necessary validation of the 
force field. 
We make a remark whose relevance will become clearer later: 
among the deformed peptides, E2f4 has the 
lowest initial energy.  

\begin{figure}[ht]
\centering
\resizebox{1\textwidth}{!}{
\includegraphics{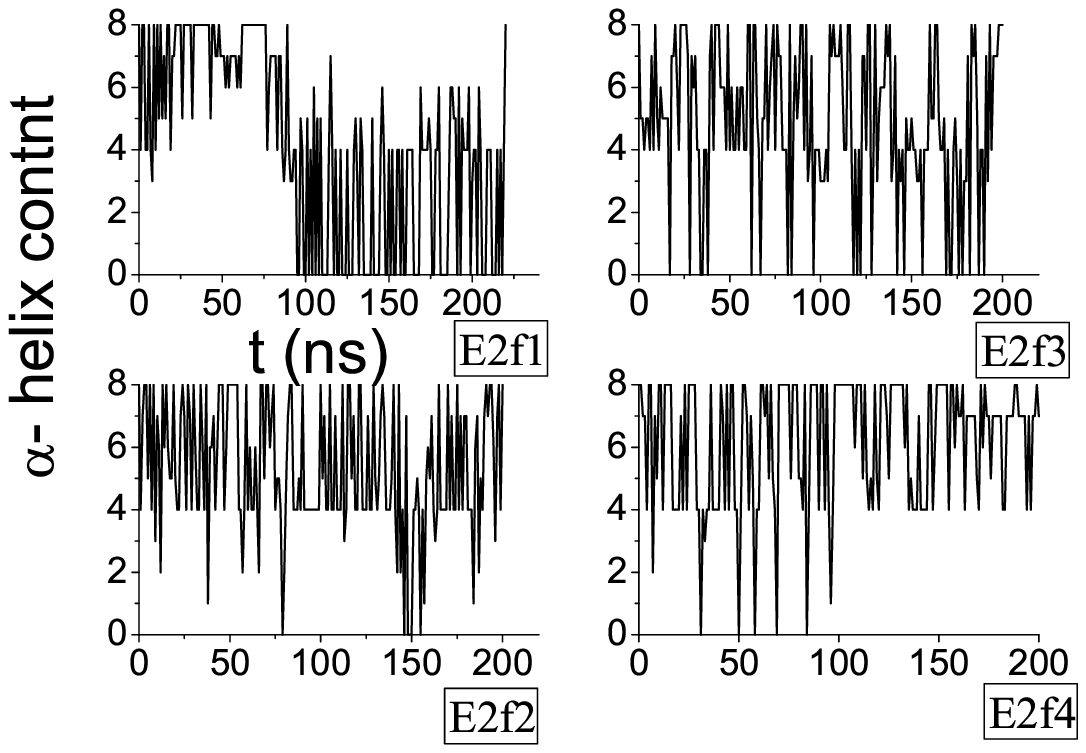}}
\caption{}
\label{f:E2-alpha-content}
\end{figure}

\noindent \textbf{H1 is unstable against modification in orientation of F1.} 
Recall that in simulations E2f1, E2f2 and E2f3, the initial relative
orientation of F1 relative H1 is changed (see Fig.\ref{f:initial_E2f}).  
Fig.\ref{f:E2-alpha-content} shows the number of residues that
constitute the $\alpha$-helix in H1 as a function of simulation time
in these simulations.  
\marginpar{Fig. 6 is here.}
Similarly, we apply the previous approaches to evaluate 95\% confidence interval of 
length between the f0 and four 
E2 simulations, which are 
$l_{f0-E2f1}=0.1006 \pm 0.0085$, $l_{f0-E2f2}=0.0933 \pm 0.0099$, 
$l_{f0-E2f3}=0.0474 \pm 0.0062$, 
and $l_{f0-E2f4}=0.003 \pm 0.0014$ respectively. 
The results put the E2f1, E2f2, and E2f3 
in an unstable class. The E2f4 and 
the f0 are in the stable class. 
It indicates that retaining F1 in the peptide but changing its orientation relative to H1 
is sufficient to destabilize 
the structure of H1. In particular, 
the modification in E2f1, which has the longest length $l_{f0-E2f1}$, has the largest effect on the destabilization of H1.


 

\smallskip\noindent \textbf{H1 is stable against modification in 
internal conformation of F1.}
In simulation E2f4 the connection between F1 and H1 is kept in its
native state but the conformation of F1 is changed by altering the
dihedral angles of Ser$^{135}$ (see Fig.\ref{f:initial_E2f}). The
simulation results are shown in the E2f4 panel of
Fig.\ref{f:E2-alpha-content}. The shortest length between the 
f0 and the E2f4, $l_{f0-E2f4}=0.003 \pm 0.0014$, suggests that there is no difference between E2f4 and f0 at 
95\% confidence level. 
The inference is that modifying 
the internal conformation of F1 does not destabilize H1.

\begin{figure}[ht]
\centering
\resizebox{1\textwidth}{!}{\includegraphics{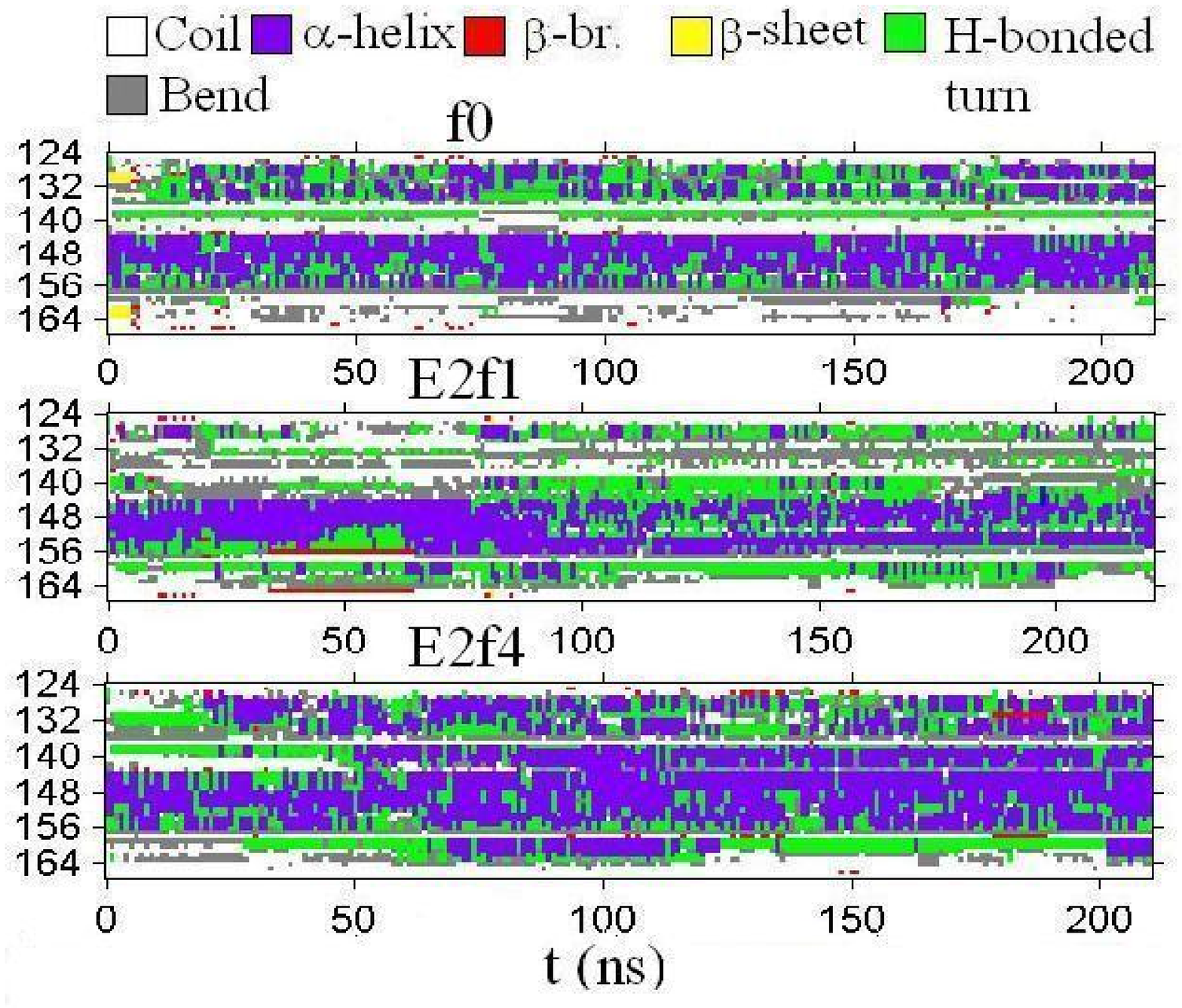}}
\caption{}
\label{f:conf_tran_E2} 
\end{figure}

\smallskip\noindent \textbf{There are new hairpin-like turns in E2f1.} 
Fig. \ref{f:conf_tran_E2} shows conformational transitions in f0,
E2f1 (representing E2f1, E2f2 and F2f3), and E2f4 during a 200 ns
simulation, where ``$\alpha$-helix'' includes the $\alpha$-like
structure $\alpha_{10}$-helix.  In all three cases, a bending site
giving a hairpin-like structure persists around residue 157 during the
full simulation. Kozin et al. in their experimental studies on the
sheep PrPC peptide in solution pointed out that this turn may form a
part of the $\beta$-sheet structure during conversion \cite{Kozin01}.
\marginpar{Fig. 7 is here.}
It is seen that f0 not only retains helical structure most of time in
the H1 region (residues 144-152) but has a tendency to form additional
helical structures around residues 126-134 as well. This last aspect
is shared by E2f4.  In contrast, E2f1 rarely has any helical structure
in the 126-134 region, has several more bends in the 132-144 region
and, after 80 ns, many more hydrogen-bonded turns in 144-152 and a
hairpin-like turn at residue 156. This hairpin-like turn is also identified to be
part of the scrapie-like   structure in the in silico study by Derreumaux \cite{Derreumaux01}. 
In addition, in the in vitro study by Megy et al., it was shown that the region 152-156, 
particularly at position 155, could be associated with the conversion \cite{Megy04}, and that in 
PB buffer a mostly  -hairpin like conformation was formed.

\begin{figure}[ht]
\centering
\resizebox{1\textwidth}{!}{\includegraphics{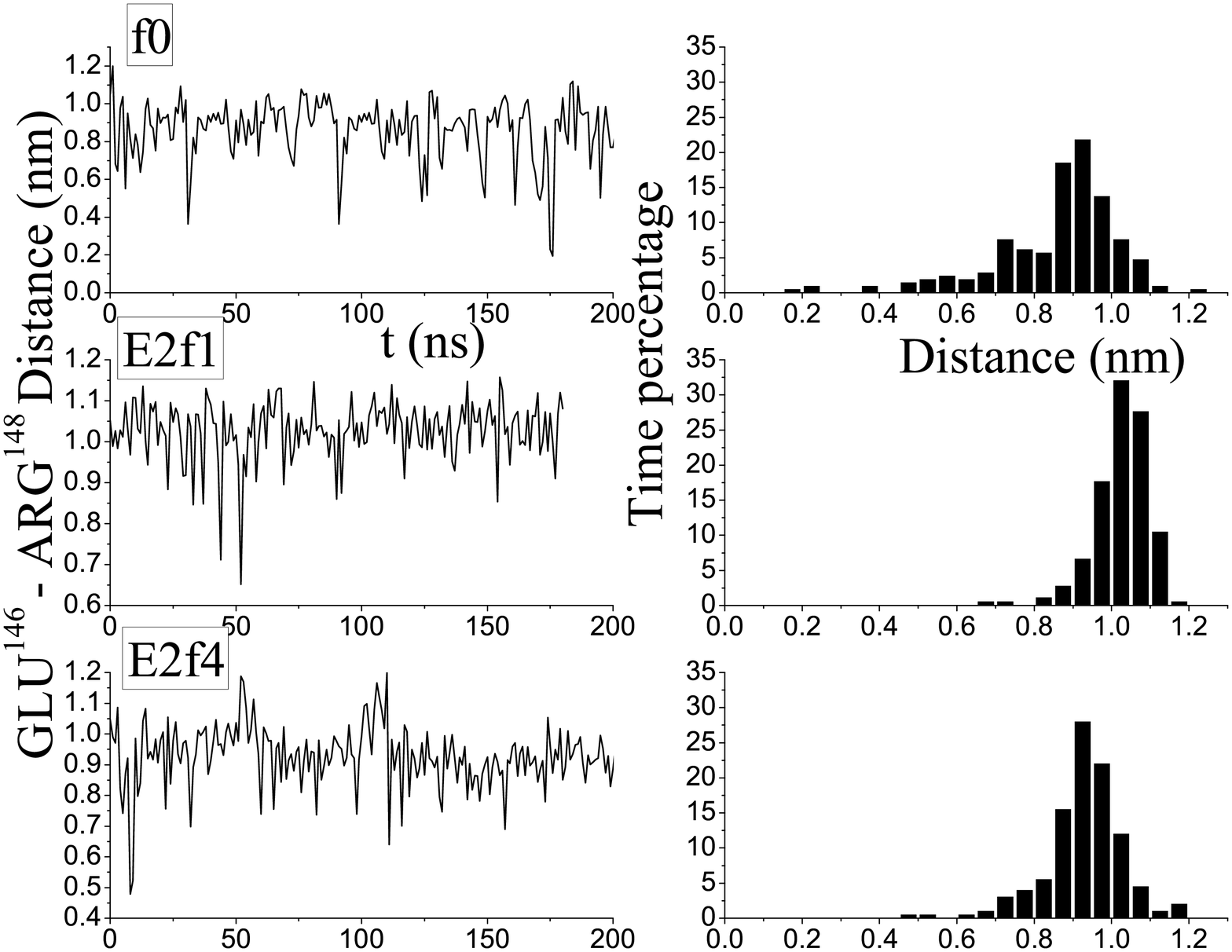}}
\caption{}
\label{f:glu-arg_a_span}
\medskip
\end{figure}

\marginpar{Fig. 8 and 9 are here.}
\smallskip\noindent \textbf{The Glu$^{152}$-Arg$^{156}$ distance 
in E2f1 is abnormally small.} 
According to Kozin et al.'s studies, the overall stability of PrP142-166 is 
associated with the three salt bridges, bonds between two oppositely charged 
groups, in the H1 region \cite{Kozin01}. These bridges connect 
the pairs of residues Glu$^{146}$-Arg$^{148}$, Asp$^{147}$-Arg$^{151}$, and
Glu$^{152}$-Arg$^{156}$. However, if one wants to stabilize an isolated H1 helical 
structure, Dima and Thirumalai suggest the first and the third salt bridges should be 
Asp$^{144}$-Arg$^{148}$ and Asp$^{148}$-Glu$^{152}$ \cite{Dima04} for an isolated H1 helix. Since 
our goal is to study overall stability of the peptide and only residues 145 and 167 of 1AG2 differ from 
the peptide they studied, we will still consider 
analyzing the three pairs suggested by Kozin et al.'s studies. 
The charges on the six residues involved in the ionic interactions
are delicately balanced and the initial distances between two oppositely charged groups in the residue pairs
for Glu$^{146}$-Arg$^{148}$, Asp$^{147}$-Arg$^{151}$, and
Glu$^{152}$-Arg$^{156}$ have spans 1.052, 0.542, and 0.936 nm,
respectively. These distances are measured from the NMR structure of the protein using the 
GROMACS tool g\_saltbr, which computes the distance between the centers of mass of two oppositely charged groups. 
Notes that g\_saltbr measures distance between center of mass of two oppositely charged groups. 
The left panel of Fig.\ref{f:glu-arg_a_span}
shows the distance of the Glu$^{146}$-Arg$^{148}$ as a
function of simulation time in the three simulations f0, E2f1 and
E2f4. The histograms in the right panel gives the percentage of simulation
time as a function of distance.  The most likely distance in the f0
and E2f4 simulations lies in the range 0.85 to 1.05 nm, whereas for
E2f1 the range is slightly greater, 0.95 to 1.1 nm.  Thus the
ionic bond in E2f1 is slightly weakened in E2f1 relative to the
two other cases, while more rigidly confined in its range.  Overall
there is no significant difference among the three cases.  The
computed distances are broadly consistent with the NMR-measured
distance of $\sim$1.05 nm for the native conformation.

\begin{figure}[h]
\centering
\resizebox{1\textwidth}{!}{\includegraphics{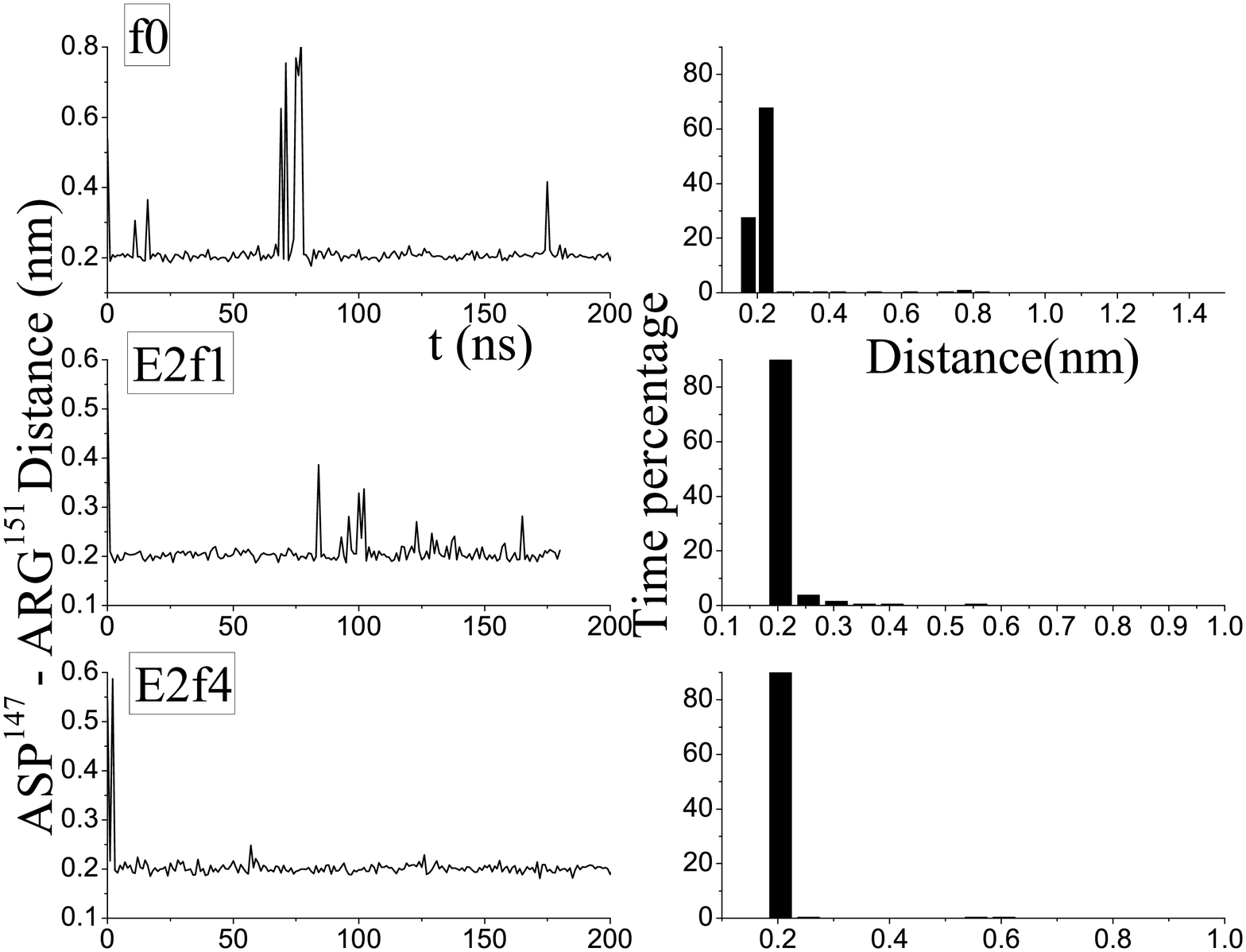}}
\caption{}
\label{f:asp-arg_span}
\end{figure}

Fig.\ref{f:asp-arg_span} (right panel) shows the span of 
the Asp$^{147}$-Arg$^{151}$ also do not vary much  
in the three simulations.  In all case the span is 
concentrated within a relatively narrow range of 0.20 to 0.25 nm. 
The computed value is however noticeably less than 
the NMR-measured distance of ~0.54 nm for the native conformation.

\begin{figure}[ht]
\centering
\resizebox{1\textwidth}{!}{\includegraphics{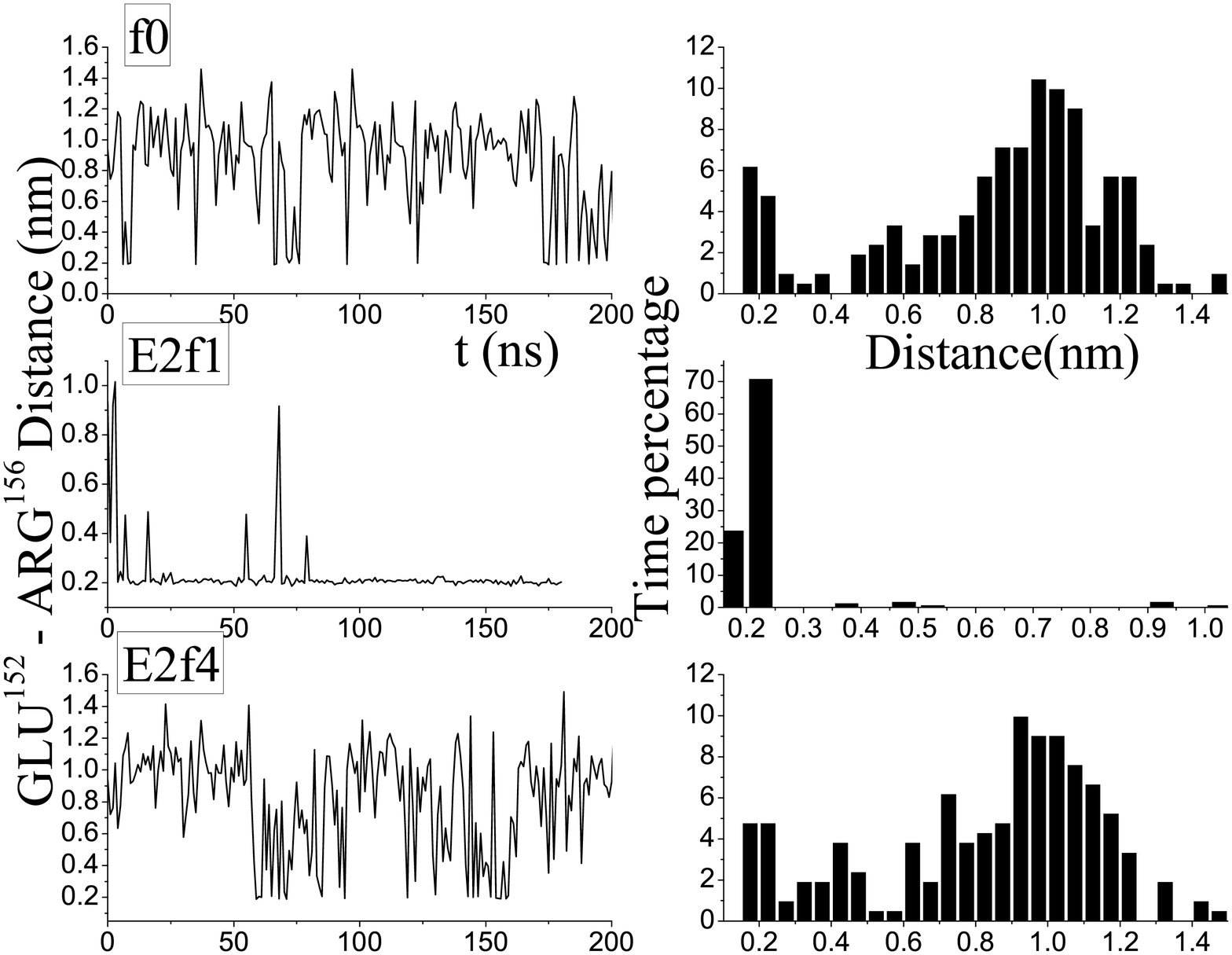}}
\caption{}
\label{f:glu-arg_b_span}
\medskip
\end{figure}

The situation shown in Fig. \ref{f:glu-arg_b_span} for the
Glu$^{152}$-Arg$^{156}$ is drastically different from the
two other pairs.  In the f0 and E2f4 simulations the distance
is mostly in the range 0.80 to 1.20 nm, and in the range 0.15 to 0.25
nm about 10\% of the time.  In sharp contrast, in the E2f1 simulation,
except for two transient periods, one at the beginning and one at
around 70 ns, the distance is less than 0.25 nm.  That the distance
settles to within a narrow range of 0.20 to 0.25 nm after 75 ns of
simulation may be correlated with the (faint) appearance of a
hairpin-like turn at residue 156 (see Fig. \ref{f:conf_tran_E2}). 
This distance is somewhat close to Megy et al. studies of sheep prion 
protein in PB buffer, which is about 0.27 nm \cite{Megy04}. Although 
our simulations does not mimic PB buffer, it seems the modification of the relative orientation of the F1 flank to 
reach a hairpin-like conformation around residue 152-156 similar to the Megy et al.'s discovery in PB buffer, 
which is believed to be likely associated with the pathogenic conversion. Namely, it may suggest the 
modification to be a required interaction for triggering conversation. It requires further investigations. 
Yet it is out of our current work's scope and will be left for future studies.
\marginpar{Fig. 10 is here.}

\smallskip\noindent \textbf{Hydrophobic does not play a major 
role in the unfolding of H1.} 
The solvent accessible surface (SAS) in a peptide gives an indication
of effect of the hydrophobic force on its structure.  In the mouse
prion 1AG2 there are nine hydrophobic residues and five non-polar
Glycines (underlined) in F1:
\underline{G}$^{124}$\underline{LGG}Y\underline{MLG}S
\underline{AM}SR\underline{PMI}H\underline{FG}N$^{142}$; five
hydrophobic residues in F2:
N$^{153}$\underline{M}YRY\underline{P}NQ\underline{V}
YYR\underline{PV}D$^{167}$; and none in H1.

\begin{figure}[ht]
\centering
\resizebox{1\textwidth}{!}{\includegraphics{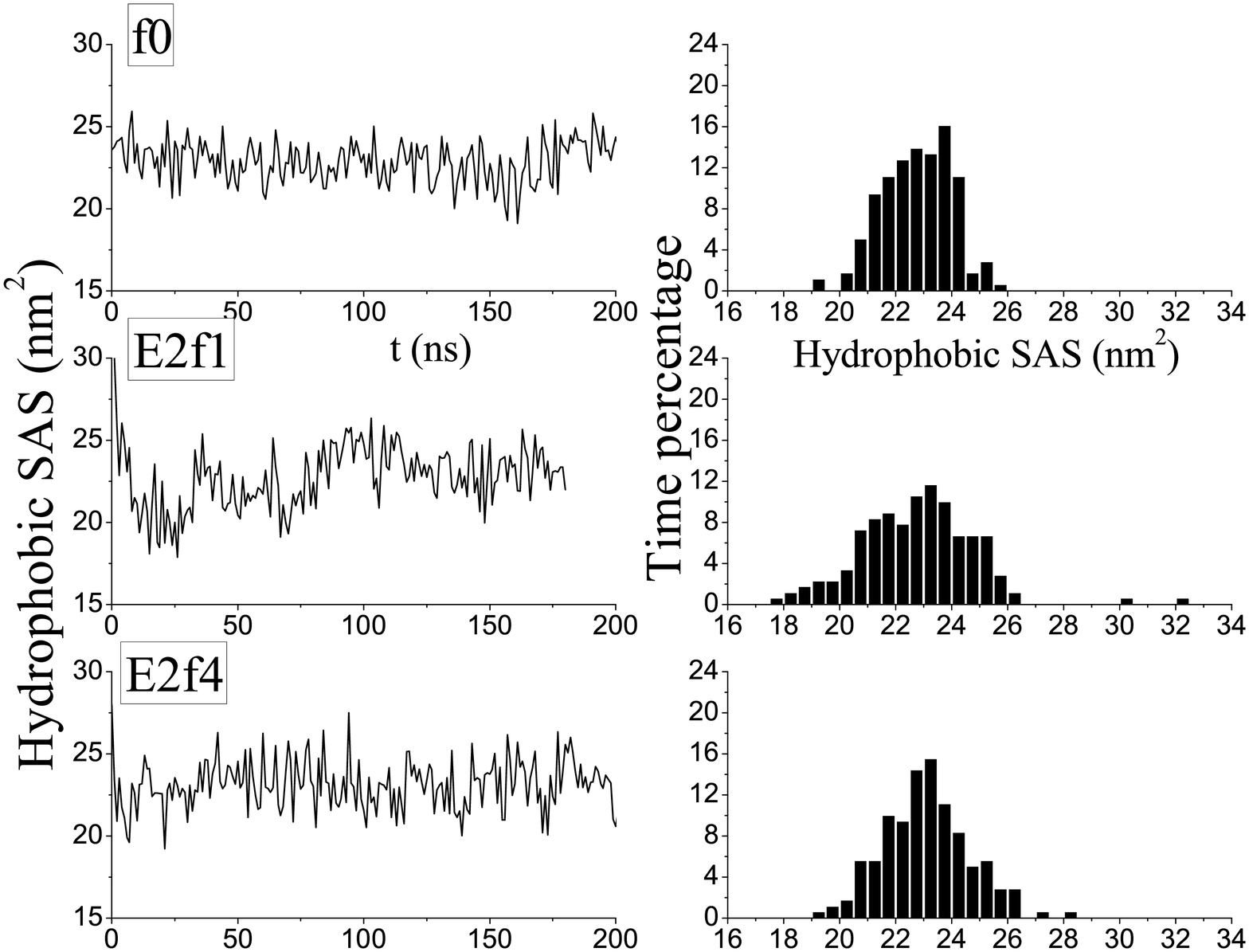}}
\caption{}
\label{f:SAS}
\end{figure}

The left panel in Fig.\ref{f:SAS} shows the SAS as a function of 
run time in the simulations f0, E2f1 and E2f4.  It is seen that in f0
the SAS mostly fluctuates around 23 nm$^{2}$ In E2f1, the
SAS shows large fluctuation - between 18 and 26 nm$^{2}$ -
during the first 100 nm of the simulation, suggesting major
conformational changes, but settles to a narrow range of 23$\pm$2
nm$^{2}$ after 100 nm$^{2}$.  In E2f4, the fluctuation is larger
then in f0, but does not have the large swings seen in E2f1.  These
fluctuations are only partly reflected in the histograms in the right
panel, which gives percentage of time against SAS.  The means and
standard deviations in SAS for the three cases are
22.9$\pm$1.3, 23.0$\pm$2.0, and 22.8$\pm$1.5 nm$^{2}$ for
f0, E2f1, and E2f4, respectively.  The larger standard deviation in
the SAS of E2F1 is caused by the conformational fluctuation in E2f1
during the early stages of the simulation.  Overall, the SAS value
does not appear to be sensitive to conformational transitions in the
S1-H1-S2 peptide. 
\marginpar{Fig. 11 is here.}

\section{Summary and Discussion}
\label{summary}
The template-assistance model attribute the pathogenesis of prion
disease to an autocatalytic process, which occurs via transient
interactions between PrPC (the mouse 1AG2 peptide) with
PrPSc. Motivated by a basic assumption of the model, that a transient
interaction may trigger a conformation conversion, we carried out MD
simulations of a simplified variant of the model.  Specifically, we
systematically imposed classes of simple alterations to the PrPC
peptide, which served as representations of the consequences of
possible transient PrPC-PrPSc interactions, and examined whether such
alterations would trigger the unfolding of the H1 region of PrPC.

The reliability of the force field used in the simulations (ff03 of
the AMBER 8 package \cite{amber}) was verified in several ways:
relative to its conformational variants, the S1-H1-S2 peptide had the
lowest simulation energy in its native conformation; if the initial
conformation of the peptide was native, then it would largely retain
its native conformation during simulation; the computed spans of the
three charged pairs of residues were in qualitative agreement with the measure
values.  The only quantitative exception was with the span of the
Asp$^{147}$-Arg$^{151}$; a computed distance of 0.20 to
0.25 nm versus a NMR measured value of ~0.54 nm.

According to various \textit{in silico} and \textit {in vitro} 
studies \cite{Eghiaian04,DeMarco07,Watzlawik06,Derreumaux01,Megy04}, 
the S1-H1-S2 peptide alone seems to be associated with the pathogenesis of the prion 
disease, and the disconnection of the H2-H3 segment from S1-H1-S2 seems to be a necessary 
condition for the latter's conversion to pathogenic form \cite{DeMarco07}.
All subsequent simulations were done on the S1-H1-S2
peptide composed of residues 124 to 163, the residues beyond F2
including the H2-H3 region - residues 164-226 - were left out.

The first series of simulations on the S1-H1-S2 peptide were designed
to show, of the two flanks F1 and F2, which would play a more
important role in preserving the $\alpha$-helical structure of H1.
Four simulations were conducted: f0, the benchmark simulation on the
entire peptide; E1f1, simulation on the peptide minus F2; E1f2,
minus F1; E1f3, minus both the F1 and F2.  The similarity analysis of these results suggested that the integrity of the helical nature of H1 
depends crucially on the presence of F1 but only weakly on the presence of F2
(Fig. \ref{f:H1-stability}).

In the second series of tests F2 was left alone and artificial
conformation alterations were made on F1 prior to simulation 
(Fig.\ref{f:initial_E2f}).  In E2f1, E2f2, and E2f3, the
relative orientations of H1 and the F1 was modified by changing the
two dihedral angles of residue Asn$^{143}$ joining F1 to H1. In E2f4
the relative orientations of H1 and the F1 were unchanged while the
internal native conformation of F1 was altered by changing the
dihedral angles of Ser$^{135}$.  It was found in the similarity analysis 
that keeping the F1-H1
angle intact was crucial to the integrity of the native H1 structure
whereas keeping the internal structure of F1 was not 
(Fig. \ref{f:E2-alpha-content}).
We remark that the present investigation is about the 
transient instability of H1 after a disturbance is made to S1-H1-S2. 
Hence the simulation that follows is not supposed to be very long;  
in the present study the duration of all simulations were 200 ns. 
Theoretically, a sufficiently long 
simulation will always bring the S1-H1-S2 peptide back to its native 
conformation, regardless of its initial state. 

Further examination of three other properties of the peptides - the
existence of hairpin-like turns, the spans of charged pairs of residues, and the
hydrophobic solvent accessible surface - showed results consistent
with our interpretation that a (specific) change to the relative F1-H1
orientation (E2f1, or E2f3, or to a lesser extent E2f2) 
would cause H1 to unravel while a change in the
internal conformation of F1 (E2f4) would not.  The spans of all three
pairs in the S1-H1-S2 peptide are similar during the f0 and
E2f4 simulations.  In the E2f1, the span of the
Glu$^{152}$-Arg$^{156}$ is reduces drastically from a native value of
about 1.0 nm to about 0.2 nm.  This change appears to correlate with
the appearance of an additional hairpin-like turn around residue 152
in the simulation of E2f1 (after 75 ns), a feature absent in the
simulations of f0 and E2f4. The SAS in the three simulations all
average to about 23 nm$^{2}$, but in the early part of the
E2f1 simulation (up to 100 ns) large fluctuation were seen, indicative
of substantial conformational changes.

The conformations of 1AG2 peptide and its subunits, the S1-H1-S2
peptide and the H2-H3 domain, all turned out to be quite robust. Our
simulations showed that neither the conformation integrity of S1-H1-S2
nor that of H2-H3 depends on the presence of the other.
Furthermore, the native H1 conformation was robust against any changes
involving F2 and changes to the internal structure of F1.  This
robustness is consistent with the fact that prion related diseases are
not easily transmitted and rarely occur spontaneously, that is, 
non-infectiously.  This may explain why, by and large, it afflicts only 
older people.  Nevertheless, there does seem to be at least one type of
vulnerability to this robustness: the helical structure of H1 is prone to
unraveling when the S1-H1-S2 peptide suffers a large change in the
relative F1-H1 orientation.

\section*{Acknowledgment}
This work is partially supported by grants 93-2811-B-008-001 
and 94-2112-M-008-013 from the 
National Science Council, Taiwan, ROC. We are grateful to the National
Center for High-performance Computing for computer time and
facilities. CYT appreciates technical help 
from HT Chen on the extraction of simulation data.

\newpage

\newpage

\section*{Figure legends}

\noindent \textbf{Figure \ref{f:1ag2-motif}}: Left: NMR structure of C-terminal of mouse PrPC; the
peptides contains three $\alpha$-helices (H), two
$\beta$-strands (S), and six surface loops (L). 
Right: The motif sequence, where the number of the leading residue 
in each motif is given.


\noindent  \textbf{Figure \ref{f:initial_E2f}}: Initial conformations of the S1-H1-S2 peptide in the E2 series of
simulations. Left: the initial conformations of E2f1, E2f2, and E2f3,
where the native values of the dihedral angles Asn$^{143}$'s are
changed. Right: the initial conformations of E2f4, where the dihedral
angles of Ser$^{135}$ are changed.

\noindent  \textbf{Figure \ref{rmsdargyr}}: Left panel plots root mean square deviation (RMSD) versus simulation time. 
Right panel plots radius of gyration.  

\noindent  \textbf{Figure \ref{f:H1-stability}}: Stability of H1 in E1 simulation.  The number of residues in the
$\alpha$-helix as a function of simulation time in three 
E1 simulations. Result of f0 is also shown for comparison.

\noindent  \textbf{Figure \ref{f:E1-family tree}}: A schematic plot of family tree for 
the f0 and the E1 simulations. The branch length is scale free.


\noindent  \textbf{Figure \ref{f:E2-alpha-content}}: Stability of H1 in E2 simulations. 
Plots show number of residues in the $\alpha$-helix  
as a function of simulation time in  E2f1, E2f2, E2f3, and E2f4.

\noindent  \textbf{Figure \ref{f:conf_tran_E2}}: Conformational transitions of the 124-167 peptide in 200 ns
simulations, with initial conformations being native (f0), E2f1, and
E2f4, respectively.  Y-axis denotes residue number and x-axis gives
simulation time.  Secondary structure are color-coded as shown at the
top of the figure, where $3_{10}$-helix is classified as
$\alpha$-helix.

\noindent  \textbf{Figure \ref{f:glu-arg_a_span}}: Variation with time in the span of the
Glu$^{146}$-Arg$^{148}$ in the f0, E2f1, and E2f4 simulations (left
panel) and percentage of time the span has a specific distance (right
panel).

\noindent  \textbf{Figure \ref{f:asp-arg_span}}: Variation with time in the span of the
Asp$^{147}$-Arg$^{151}$ in the f0, E2f1, and E2f4 simulations (left
panel) and percentage of time the span has a specific distance (right
panel).

\noindent  \textbf{Figure \ref{f:glu-arg_b_span}}: Variation with time in the span of the Glu$^{152}$-Arg$^{156}$ 
in the f0, E2f1, and E2f4 simulations (left panel) and 
percentage of time the span has a specific distance (right panel).

\noindent  \textbf{Figure \ref{f:SAS}}: Variation with time in the hydrophobic SAS 
in the f0, E2f1, and E2f4 simulations (left panel) and 
percentage of time the SAS has a specific value (right panel).

\newpage
\section*{Figures}

\newpage


\newpage

\newpage

\newpage

\newpage

\newpage


\newpage

\newpage

\newpage

\newpage

\newpage

\newpage

\begin{thebibliography}{1}

\bibitem{Aguzzi08a} Aguzzi A, Sigurdson C, Heikenwaelder M (2008a) 
Molecular Mechanisms of Prion Pathogenesis. Annu. Rev. Pathol. 3:11-40. 

\bibitem{Aguzzi08b} Aguzzi A, Baumann F, Bremer J (2008b) The Prion's Elusive Reason for Being. 
Annu. Rev. Neurosci. 31:439-477.

\bibitem{Barducci06} Barducci A, Chelli R, Procacci P, Schettino V, Gervasio FL, 
Parrinello M (2006) Metadynamics simulation of prion protein: $\beta$-structure stability and the early stages of misfolding. 
2006. J. Am. Chem. Soc. 128:2705-2710.

\bibitem{GROMACS2}Berendsen HJC, van der Spoel D, van
Drunen R (1995) GROMACS: A message-passing parallel molecular dynamics
implementation. Comp. Phys. Comm. 91:43--56 

\bibitem{Calzolai03} Calzolai L and Zahn R (2003) Influence of pH on NMR
structure and stability of the human prion protein blobular
domain. J. Biol. Chem. 278:35592-35596.

\bibitem{amber} Case DA, Cheatham TE, Darden T, Gohlke H, Luo R, Merz Jr. KM, 
Onufriev A, Simmerling S, Wang B, Woods R (2005) The AMBER
biomolecular simulation programs. J. Computat. Chem. 26:1668-1688.

\bibitem{Castilla06} Castilla J, Sa\'a P, Hetz C, Soto C (2006) In vitro generation 
of infectious scrapie prions. Cell 121:195-206.

\bibitem{pymol} DeLano WL (2002) PyMOL molecular graphics system (DeLano
Scientific, San Carlos, CA, USA.). http://www.pymol.org/. 

\bibitem{DeMarco04} DeMarco ML and Daggett V (2004) From conversion to
aggregation: Protofibril formation of the prion
protein. Proc. Natl. Acad. Sci. USA. 101:2293-2298.

\bibitem{DeMarco07} DeMarco ML and Daggett V (2007) Molecular mechanism for low
pH triggered misfolding of the human prion protein. Biochem. 46:3045-3054.

\bibitem{Derreumaux01} Derreumaux P (2001) Evidence that the 127-164 region of prion proteins has 
two equi-energetics conformations with $\beta$ or $\alpha$ features. Biophys. J. 81:1657-1665.

\bibitem{Dima04} Dima RI and Thirumalai D (2004) Probing the instabilities in the dynamics of helical 
fragments from mouse PrP$^{\textnormal{C}}$. Proc. Natl. Acad. Sci. USA. 101:15335-15340.


\bibitem{Eghiaian04} Eghiaian F, Grosclaude J, Lesceu S, Debey P,
Doublet B, Treguer E, Rezaei H, Knossow M (2004) Insight into the
PrPC$\rightarrow$PrPSc 
conversion from the structures of antibody-bound ovine prion
scrapie-susceptibility variants. Proc. Natl. Acad. Sci. USA.
101:10254-10259.

\bibitem{deepview} Guex N and Peitsch MC (1997) SWISS-MODEL and the
Swiss-PdbViewer: An environment for comparative protein
modeling. Electrophoresis. 18:2714-2723.

\bibitem{Guilbert00} Guilbert C, Ricard F, Smith JC (2000) Dynamic
simulation of the mouse prion protein. Biopolymers. 54:406-415.

\bibitem{Harris99} Harris DA (1999) Cellular biology of prion diseases. 
Clin. Microbiol. Rev. 12:429-444.

\bibitem{Horiuchi99} Horiuchi M and Caughey B (1999) Prion protein
interconversions and the transmissible spongiform
encephalopathies. Structure Fold. Des. 7:R231-R240.

\bibitem{Jackson00} Jackson GS and Clarke AR (2000) Mammalian prion proteins. 
Curr. Opin. Struct. Biol. 10:69-74.


\bibitem{dssp} Kabsch W and Sander C (1983) Dictionary of protein secondary
structure: pattern recognition of hydrogen bond and geometrical
features. Biopolymers. 22:2577-2637.

\bibitem{Kozin01} Kozin SA, Bertho G, Mazur AK, Rabesona H, Girault JP, 
Haertie T, Takahashi M, Debey P, and Hoa GHB (2001) Sheep prion protein
synthetic peptide spanning helix 1 and $\beta$- strand 2 (Residue
142-166) shows $\beta$-hairpin structure in solution. J. Biol. Chem. 
49:46364-46370.

\bibitem{Levy01} Levy Y, Hanan E, Solomon B, Becker OM (2001) Helix-coil
transition of PrP106-126: Molecular dynamic study. Proteins. 45:382-396.

\bibitem{Levy02} Levy Y and Becker OM (2002) Conformational polymorphism of
wild-type and mutant prion proteins: Energy landscape
analysis. Proteins. 47:458-468.

\bibitem{GROMACS3}Lindahl E, Hess B, van der Spoel D (2001) Gromacs 3.0: A package for molecular simulation
and trajectory analysis. J. Mol. Mod. 7:306-317.

\bibitem{Malolepsza05} Malolepsza E, Boniecki M, Kolinski A, Piela L (2005) 
Theoretical model of prion propagation: A misfolded protein
induces misfolding.  Proc. Natl. Acad. Sci. USA. 102:7835-7840.

\bibitem{Megy04} Megy S, Bertho G, Kozin SA, Deby P, Hoa GHB, 
Girault J-O (2004) Possible role of region 152-156 in the structural duality of a peptide fragment 
sheep prion protein. Protein Sci. 13:3151-3160.

\bibitem{Morrissey99} Morrissey MP and Shakhnovich EI (1999) Evidence for the
role of PrP$^{C}$ helix 1 in the hydrophilic seeding of prion aggregates.
Proc. Natl. Acad. Sci. USA. 96:11293-11298.

\bibitem{Muramoto96} Muramoto T, Scott M, Cohen FE, Prusiner SB (1996) 
Recombinant scrapie-like prion protein of 106 amino
acids. Proc. Natl. Acad. Sci. USA. 93:15457-15462.

\bibitem{Onufriev00} Onufriev A, Bashford D, Case DA (2000)
Modification of the generalized Born model suitable for
macromolecules. J. Phys. Chem. B 104:3712-3720.


\bibitem{Pan93} Pan KM, Baldwin M, Nguyen J, Gasset M, Serban A, Groth D
, Mehlhorn I, Huang Z, Fletterick RJ, Cohen FE (1993) Conversion of
$\alpha$-helices into $\beta$-sheets features in the formation of
the scrapie prion proteins. Proc. Natl. Acad. Sci. USA.
90:10962-10966.

\bibitem{Prusiner98} Prusiner SB (1998) Prions. Proc. Natl. Acad. Sci. USA.
95:13363-13383.

\bibitem{Riek96} Riek R, Hornemann S, Wider G, Billeter M, Glockshuber R, 
Wuthrich K (1996) NMR structure of the mouse prion protein domain
PrP. Nature. 382:180-182.

\bibitem{Ryckaert97} Ryckaert JP, Ciccotti G, Berendsen HJC (1997)
Numerical integration of the cartesian equations of motion of a system with constraints:
Molecular dynamics of n-alkanes. J. Comput. Phys. 23:327-341.


\bibitem{Santini03} Santini S, Calude J-B, Audic S, Derreumaux P (2003) Impact of tail and mutations 
G131V and M129V on prion protein flexibility. Proteins. 51:258-265.

\bibitem{Santini04} Santini S and Derreumaux P (2004) Helix H1 of the prion protein is rather stable 
against enviormental perturbations: molecular dynamics of mutation and deletion variants of PrP (90-231). 
Cell. Mol. Life Sci. 61:951-960.

\bibitem{Sharman98} Sharman GJ, Kenward N, Williams HE, Landon M,
Mayer RJ, and Searle MS (1998) Prion protein fragments spanning helix 1 and
both strands of $\beta$-sheet (residues 125-170) show evidence for
predominantly helical propensity by CD and NMR. Folding \& Design. 3:313-320.

\bibitem{Tompa02} Tompa P, Tusnady GE, Simon I (2002) The role of
dimerization in prion replication. Biophys. J. 82:1711-1718.

\bibitem{Watzlawik06} Watzlawik J, Sokra L, Frense D, Griesinger C, 
Zweckstetter M, Schulz-Schaeffer WJ, Kramer ML (2006) Prion protein helix 1 promotes aggregation but 
is not converted into $\beta$-sheet. J. Bio. Chem. 281:30242-30250.

\bibitem{GROMACS} van der Spoel D, Lindahl E, Hess B, van Buuren AR,
Apol E, Meulenhoff PJ, Tieleman DP, Sijbers ALTM, Feenstra KA, van
Drunen R, Berendsen HJC (2004) Gromacs user manual version
3.2. http://www.gromacs.org/ 


\bibitem{Zou02} Zou WQ and Cashman NR (2002) Acidic pH and detergents
enhance in vitro conversion of human brain PrPC to
PrPSc-like form. J. Biol. Chem. 277:43492-43947.

\bibitem{Zeigler03} Ziegler J, Sticht H, Marx UC, Muller W, Rosch P, 
Schwarzinger S (2003) CD and NMR Studies of Prion Protein (PrP) Helix
1. J. Biol. Chem. 278:50175-50181.






\end{thebibliography}
\end{document}